# Broad Distribution of Local I/Br Ratio in Illuminated Mixed Halide Perovskite Films Revealed by Correlative X-ray Diffraction and Photoluminescence

AUTHOR NAMES Klara Suchan[1], Justus Just[2], Pascal Becker[3,4], Carolin Rehermann[5], Aboma Merdasa[5], Roland Mainz[4], Ivan G. Scheblykin*[1] and Eva L. Unger*[1,5]

AUTHOR ADDRESS [1]Division of Chemical Physics and NanoLund, Lund University, Box 124, 22100 Lund, Sweden, [2]MAX IV Laboratory, Lund University, PO Box 118, SE-22100 Lund, Sweden, [3]Helmholtz-Zentrum Berlin für Materialien und Energie GmbH, Structure and Dynamics of Energy Materials, Hahn-Meitner-Platz 1, 14109 Berlin, Germany, [4]Helmholtz-Zentrum Berlin für Materialien und Energie GmbH, Microstructure and Residual Stress Analysis, Albert-Einstein-Straße 15, D-12489 Berlin, Germany, [5]Helmholtz-Zentrum Berlin für Materialien und Energie GmbH, HySPRINT Innovation Lab: Hybrid Materials Formation and Scaling, Kekuléstraße 5, 12489 Berlin, Germany

AUTHOR INFORMATION

Corresponding Authors *E-Mail address: eva.unger@helmholtz-berlin.de, ivan.scheblykin@chemphys.lu.se

ABSTRACT Photo-induced phase-segregation in mixed halide perovskite MAPb($Br_xI_{1-x}$)$_3$ is investigated in the full compositional range by correlative X-ray diffraction and photoluminescence experiments. The compositional redistribution prior and upon illumination is quantitatively derived from detailed X-ray diffraction analysis and is shown to depend on the sample's average composition. As previously reported, the luminescence peak energy evolving during illumination is almost independent of the average sample composition for all mixed halide samples, suggesting that similar low-energy emitting sites are formed in all samples. X-ray diffraction analysis reveals that the majority of the material exhibits a broad compositional distribution with locally varying iodide to bromide ratios.

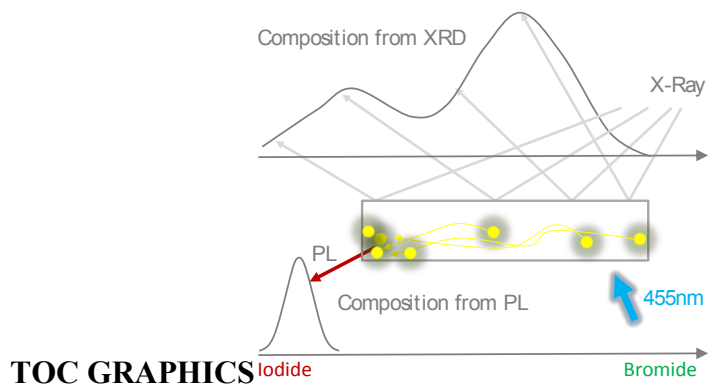

**TOC GRAPHICS**

MAIN TEXT Metal-halide perovskite (MHP) solar cells have attracted increasing attention due to their rapid increase in power conversion efficiency (PCE) to now above 25.2 %, close to values of established technologies such as crystalline silicon solar cells.[1] Tandem solar cells, combining perovskite with existing silicon technology have recently reached certified power conversion efficiencies of 28 %,[1] thus proving to be a promising route of decreasing the overall cost of solar energy conversion technology. The ideal bandgap of the top cell absorber for tandems based on silicon or Cu(In,Ga)Se$_2$ as bottom cells lies between 1.65 eV and 1.8 eV.[2–6] By changing the nominal average composition <x> of the halides, iodide and bromide in APb(Br$_x$I$_{1-x}$)$_3$ (with A =MA, FA, Cs or a mixture thereof),[7] the bandgap of perovskite absorbers can be continuously tuned from 1.45 eV to 2.4 eV.[7,8] Note that we herein refer to the nominal average composition of the sample as <x>, derived from the halide ratio in the precursor solution. The local composition x within the sample may differ from this significantly due to compositional inhomogeneities on the nano- to micrometer scale.

A limitation to the bandgap tunability by mixing halides is their reported light-induced phase segregation.[9] A red-shift of the photoluminescence (PL) emission energy during illumination indicates the formation of low energy emission sites linked to the segregation of the halide ions into I-rich and Br-rich local domains due to halide ion migration. This phenomenon has been observed in all in-situ XRD studies under illumination of higher bandgap mixed bromide/iodide perovskite samples investigated so far.[9–15] This red-shift is rationalized by charge carriers created within the sample which are funneled to I-rich small bandgap domains where they recombine.[9] This compositional instability coincides with an increased loss of open circuit voltage (V$_{OC}$)[7,16] compared to the band-gap of the material and a miscibility gap observed in powders[17]. It was proposed, that the formation of I-rich domains, results in a decrease of V$_{OC}$ and subsequently limits device efficiencies of higher bandgap mixed halide perovskites.[16,18,19]

For methylammonium-based mixed halide perovskite samples, MAPb(Br$_x$I$_{1-x}$)$_3$, the reported PL peak energies evolving upon illumination for a bromide content x > 0.1 lie between 1.65 eV and 1.75 eV.[12,16] The PL thus resembles that of band-to-band emission from samples with a composition of <x> = 0.1-0.2. This universality of the PL peak energy led to the speculation that there may be preferred stable compositions of x = 0.2 and x = 0.8 into which the material locally segregates, independent on the samples average composition.[9,12] These compositions coincide with theoretically predicted thermodynamically favored compositions which are found to be the limits of miscibility in powder samples of MAPb($Br_xI_{1-x}$)$_3$ prepared under thermodynamic reaction control.[10,17,20]

So far, few studies have used X-ray diffraction (XRD) measurements to investigate the phase segregation in APb($Br_xI_{1-x}$)$_3$, however only a single composition or very few compositions within a limited composition range (<x> = 0.2 and 0.6) were considered.[9,12,13,15,17,21,22] While some studies suggest a split into a bromide-rich phase and an iodide-rich phase, others postulated merely a strain related peak broadening without an observable phase segregation.[9,12,13,15,21] Furthermore, there is a large discrepancy in the estimation of the fraction of the material in the low energy phase, which was reported to be between 2% and 23%.[12,23] [9] These reports may not necessarily contradict each other but rather reflect a dependence of the compositional distribution upon illumination on the sample's average composition, which we will demonstrate by our study below.

Here, we performed correlative XRD and PL measurements, to shed light on the structural changes upon photo-induced phase-segregation on a systematic series of MAPb($Br_xI_{1-x}$)$_3$ perovskite samples for nominal sample compositions <x> ranging from 0 to 1. To minimize potential pre-existing phase segregation in samples due to synthesis or exposure to ambient light prior to the experiment,[27,28] samples were heated to 90 °C in dark conditions before the measurement. The phase-segregated state induced upon illumination by blue LED light was analyzed using X-ray diffraction and PL. While both PL and XRD can give compositional information, their sensitivity is fundamentally different. While PL is very sensitive to and often completely dominated by the lowest energetic states present, XRD gives a quantitative estimate of all crystalline phases in the sample, provided large enough domains exist. By comparing and contrasting the results from both measurement techniques, we gain necessary information on the compositional distribution after phase segregation, not accessible via PL alone, providing unprecedented insight into the composition-dependence of light induced phase segregation and equilibrium compositional distribution established upon illumination.

Simultaneous X-ray diffraction and photoluminescence measurements on samples before and upon exposure to light were performed under inert atmosphere in a custom-built setup (Figure 1a). For experimental details see Supplemental Information. Upon illumination the XRD pattern, shown exemplary for the sample with <x> = 0.6 (Figure 1b), exhibits a significant reduction in the peak height accompanied by peak broadening at all visible reflexes of the cubic $Pm\bar{3}m$[24] crystal structure. No additional diffraction peaks were observed. Furthermore, the integral peak intensity, which is approximately proportional to the amount of crystalline material, remains constant. We thus exclude the degradation of the material into non-perovskite phases. As PL cannot be measured in the absence of light, the PL spectrum of the first second is compared to the one upon further illumination, at which both the XRD signal and the PL signal have stabilized. The original PL lies at 1.9 eV, however already within the first second a low energy peak at 1.7 eV has evolved. Subsequently the red-shifted PL is increasing as reported previously (Figure 1b). [9]

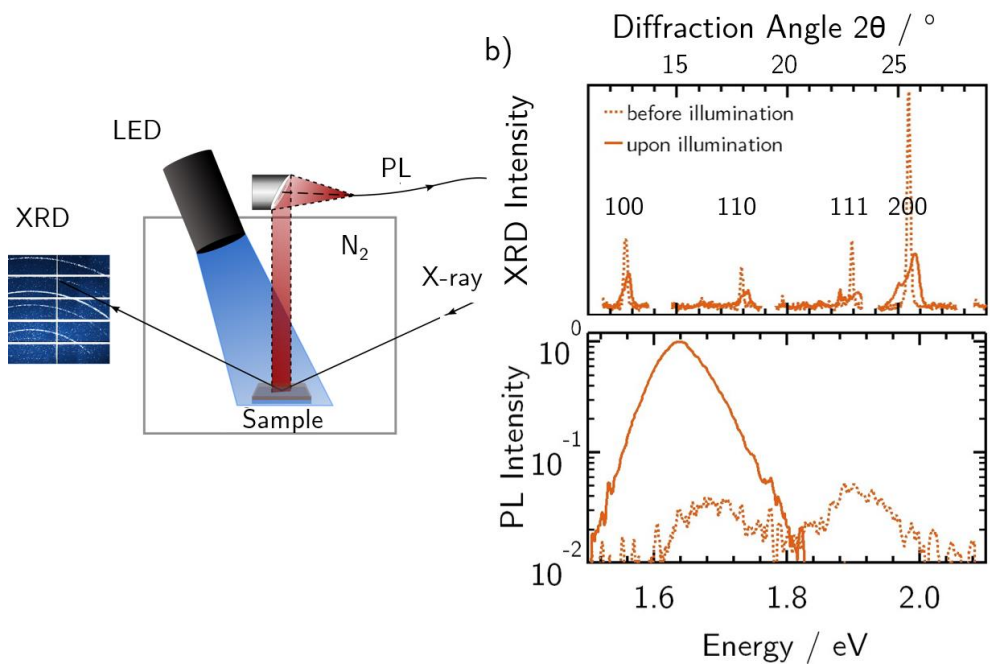

**Figure 1:** a) Schematic drawing of the measurement setup used to study the change in the structural and optical properties of $MAPb(Br_xI_{1-x})_3$ before and upon illumination. b) X-ray diffraction (XRD) patterns and PL spectra of the sample $\langle x \rangle = 0.6$, before and upon illumination (XRD wavelength: 1.3401 Å). XRD measurements where performed using $GaK_\alpha$ radiation (9.251 keV).

The XRD peaks corresponding to the 200 reflex for the cubic $Pm\bar{3}m$ and the 220/004 reflex for the tetragonal P4mm perovskite phase are shown in Figure 2a before and after illumination for all $\langle x \rangle$. The pure I and Br perovskites ($\langle x \rangle = 0$ and $\langle x \rangle = 1$) do not exhibit an observable change in XRD, during illumination (Figure S2). We thus conclude, that neither the visible illumination nor X-ray exposure affect the structural properties of these pure halide samples under our experimental conditions. Please note that we deliberately did not correct for the $K_{\alpha 1}/K_{\alpha 2}$ peak split, which is apparent in the experimental data of the iodide and bromide samples.

The samples with compositions close to the pure compositions (<x> ≥0.8 and <x> ≤0.2) exhibit a minor decrease in the peak height after illumination, accompanied by composition-dependent broadening and a peak shift to higher (lower) angles for <x>≥0.8 (<x> ≤0.2). Such a shift of the XRD diffraction peak of bromide rich samples to higher angles has previously been reported and explained by an increase of inhomogeneous strain, distortion and decrease of coherence volume.[12] Changes or shifts in the XRD peaks, as we observe for the I-rich samples, have previously not been reported. This is probably due to the fact that it was assumed that iodide-rich samples would not exhibit any structural changes as no changes was observed in PL spectra. Here we clearly show that sample with <x>=0.1, previously assumed to be photo-stable[9], indeed exhibits a slight change of the position and broadening of the diffraction peak upon illumination. This composition exhibits a PL enhancement upon light soaking as also observed for the pure I-perovskites[25] but no change in the PL spectral position. Thus, the structural change observed in the XRD pattern for <x> = 0.1 does affect the PL. Observation of the structural changes upon illumination hence indicate a compositional instability beyond the instability observed in PL.

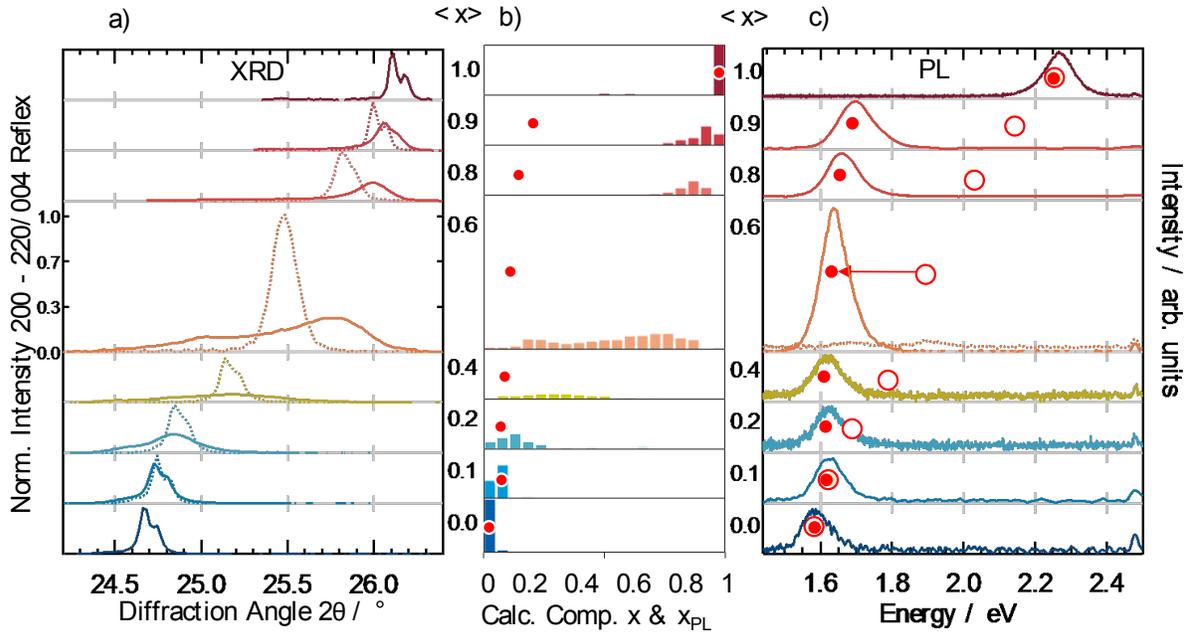

**Figure 2.** a) X-ray diffraction pattern of the 200 (cubic) or 220/004 (tetragonal) reflex for different average compositions <x> (XRD wavelength 1.3401 Å). The dashed lines indicate the initial pattern and the solid lines the phase-segregated pattern after illumination. b) Histograms of the compositional (x) distribution in the samples after illumination as calculated from XRD (See Fig 4). The composition $x_{PL}$ calculated from the final PL peak energy is shown by solid red circles. c) PL spectra of the illuminated films for different <x>. The PL of the phase segregated film is red-shifted compared to the initial film, as shown exemplarily for the sample <x>=0.6. The initial (open circle) and final (solid circle) positions of the PL spectrum are shown for all <x>, arrows demonstrate the PL shift after illumination. The dependence of the initial PL position on <x> is used as the calibration curve to calculate $x_{PL}$ from x (shown in b).

The XRD pattern of the composition <x> = 0.6 exhibits two broad peaks shifted to lower and higher angles compared to the initial peak. Such a split in XRD peaks upon illumination has been reported for samples of <x> = 0.4 and 0.6. was explained by a segregation into two distinct compositions.[9,22] For <x> = 0.6, we additionally observe broadening of each diffraction peak. In our sample set, the composition of <x> = 0.4 shows a similarly broad intensity distribution however lacking the segregation into two peaks. In the presented composition dependent XRD data, we were hence able to reproduce all qualitatively different light-induced changes in XRD patterns reported previously. The previous disagreement in the literature XRD data can thus be reconciled by a dependence of the structural changes on the sample's average composition.

We chose to show histograms of the phase-distribution of samples with different halide compositions upon illumination in direct comparison with the experiment XRD patterns in Figure 2b. Exact details of how these were derived will be discussed below. In contrast to the XRD patterns, the spectral position of the PL peak after illumination is quite similar for all compositions (Figure 2c), in agreement with literature.[9] We note only a slight systematic blue shift upon increasing the Br content.

Our results demonstrate the complementarity in experimental insight that can be gained from XRD and PL data. The formation of I-rich domains is clearly indicated by the emergence of a lower energy PL feature that is similar in energy for all compositions with <x> > 0.1. XRD results however prove how vastly different the actual phase distribution becomes for samples under illumination for various bromide to iodide ratios. The XRD pattern of the sample with <x> = 0.6 shows an obvious split indicating an iodide-rich and a bromide rich phase distribution. The patterns of all other samples only show a peak shift and broadening, which has previously been interpreted as an increase in strain rather than a phase segregation.[12] To investigate this further the peak broadening is analyzed.

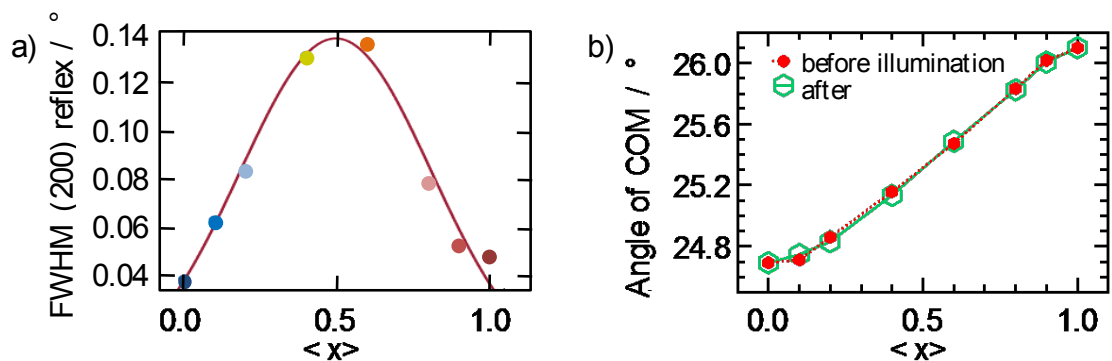

**Figure 3**: a) Comparison of the FWHM of the 200 reflex for different <x> prior to illumination shown to follow a gaussian distribution as indicated by the fit (red line). c) Center of mass of the pattern before and after illumination depending on composition.

Already in the initial state before illumination the peak broadening is found to be composition dependent following a gaussian distribution (Fig. 3a). The pure I or Br samples show quite narrow peaks while the width is increasing upon halogen substitution reaching the largest broadening at <x> = 0.5. A comparison of the Williamson-Hall analysis for <x> = 0 and <x> = 0.6 indicates, that the increased peak width for samples prior illumination originates primarily from a lattice distance inhomogeneity in the mixed halide crystal[26] (Fig. S4a) causing a maximum in the FWHM of XRD peaks at equal I and Br content. In previous reports showing maximal FWHM broadening of XRD peaks at higher Br compositions, samples likely exhibit more pronounced pre-existing phase-segregation,[27,28] which we here avoided by heating samples to 90 °C in dark conditions immediately before the measurement.

The mixed samples show an additional broadening upon illumination. Fig. S5 illustrates, that even the sample <x>=0.6 cannot be described by merely a split into two distinct compositions, but in this case each peak is additionally broadened. The Williamson-Hall analysis (Fig. S4b) reveals that this increase in broadening can be rationalized by an increase of the lattice distance inhomogeneity rather than a decrease of the domain size. This indicates that the light induced phase segregation does not result in a decrease of the crystallite size. However, any lattice distance inhomogeneity will look indistinguishable in a Williamson-Hall plot. Thus, a closer look at the peak shape can help to disentangle whether this inhomogeneity is strain related or due to compositional inhomogeneity.

Micro strain commonly leads to a homogeneous (Gaussian) peak broadening. The strained peak should thus resemble a convolution of the original peak shape (Pseudo Voigt) and a Gaussian.[29] On the contrary, a broadening due to compositional inhomogeneity does not have this constraint. In the samples with compositions closer to pure I and pure Br ($x < 0.4$ and $x > 0.6$), illumination induces a peak shift accompanied by peak broadening, which might look as due to an increased strain in the sample. However, a more detailed look at all patterns shows, that even though the peak maximum is shifted significantly (e.g. in the composition $<x>=0.8$ the peak shifts by 0.2 degrees), the arithmetic mean of all points on the horizontal axis weighted by their individual intensity (further referred to as center of mass, discussed in SI) of all patterns stays constant (Figure 3b). This is due to a broad distribution of minor scattering intensity at lower angles (24.5-25°). The constant center of mass of the patterns shows that despite of the, for some compositions, large shift of the peak maximum, the net-strain in the sample stays constant during illumination. Any increase in compressive or tensile strain in the sample would thus have to be exactly compensated. This would lead to a symmetric Gaussian broadening around the position of the original peak (further discussed in SI).[29] The measured patterns, as shown in Figure 2a, however are asymmetric with respect to the original peak position. This allows for the assumption that during the phase segregation there is no relevant increase in strain. Instead, all changes in the XRD patterns are consistent with phase segregation into a broad distribution of compositions x. If the broadening is induced by compositional inhomogeneity, the average composition <x> and thus the center of mass of the patterns must stay constant consistent with our observation. A variation of the local compositions within the film will lead to a multitude of narrow peaks at individual diffraction

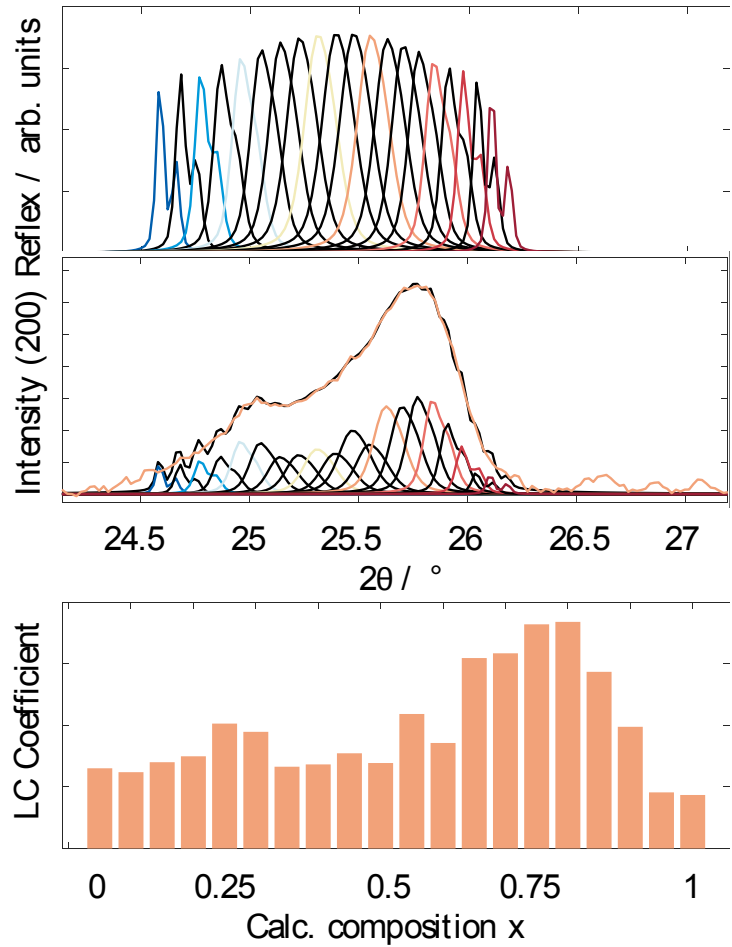

**Figure 4** a) The X-ray diffraction patterns of the 20 compositions used to fit the patterns of the phase segregated films with their linear combinations. b) The diffraction pattern of the phase segregated film of composition <x>=0.6 (orange), the fit (black) and the contributing pattern (color according to Fig 2). c) The resulting linear combination coefficients are shown in a phase distribution histogram.

angles. The resultant diffraction pattern can hence be interpreted as a linear combination of XRD peaks from different (local) compositions. With this approach, both a peak shift and broadening as well as the peak splitting can be described consistently.

We chose to split the whole range of x from 0 to 1 into 20 compositions (Figure 4a) and to fit all XRD patterns with a linear combination of these 20 patterns. Figure 4b shows the resulting fit (further described in the SI) exemplary for $<x> = 0.6$. The coefficients of the linear combination of the 20 compositions provide a quantitative measure of the distribution of phases in the phase segregated sample (Figure 4c). The resulting estimation of the compositional distribution x is shown for all samples investigated in Figure 2b.

We found that the very I-rich compositions (x < 0.2), often assumed to be photo-stable, also show a change in their compositional distribution upon illumination, if yet narrow. However, in the case of $<x>$=0.4 and 0.6 almost the entire range of compositions from $0.1 < x > 0.8$ can be found upon illumination. Br-rich samples exhibit phase-segregation into a narrow high-energy phase, with a very broad tail of low energy phases containing almost all possible compositions down to x=0.1.

The composition of the emissive phase can be calculated from PL ($x_{PL}$) using the dependence of PL on $<x>$ for the well-mixed samples before light-induced segregation as the calibration curve (Figure S3). By comparing the distribution of compositions x and $x_{PL}$ it becomes obvious that for samples with $<x> > 0.2$, the PL originates from a composition with an iodide content of approximately x = 0.2 stemming from few iodide-rich phases at the end of the low energy tail in XRD. In the case of $<x> = 0.6$ for example, the PL originates from a composition of 10% Br ($x_{PL}$ = 0.1) adopted only by 1% of the sample volume. Even the material giving the maximum of the low-energy XRD peak has a much higher bromide content (30 % bromide). The <x>=0.8 sample exhibits a very narrow XRD peak with maximum at 90% Br (x=0.9), without a clear low energy peak. Nonetheless, the PL emission is coming from 15% Br ($x_{PL}$=0.15).

Although, both PL and XRD can give insights into the compositional phases present in a mixed halide perovskite, their sensitivity is based on completely different selection criteria. Due to effective charge transport, PL selects the lowest accessible energy states of the overall distribution. It therefore gives information about the lowest energetic phase present in the material which agrees with the hardly visible tail in the distribution of x obtained from XRD. However PL cannot give information about the amount of material possessing those energy states, as the PL quantum yield is highly dependent on charger carrier funneling, light soaking and local defect concentrations.[25,30,31] This means that using PL as a sole source of information might greatly overestimate the amount of low-energy composition.

Our results show that the final composition distribution of the light-induced phase segregated films of MAPb($Br_xI_{1-x}$)$_3$ is highly dependent on the average composition <x> of the film. The picture that all samples will segregate into two, universally favored, low and high energy compositions[12,20] was not confirmed experimentally. Even if only one peak is visible in the XRD pattern, it is more accurately described by a compositional distribution of phases rather than one strained phase. Elucidating the actual structural composition of the phase segregated state, we show that all samples contain low-energy I-rich phases down to x = 0.1. The energy of these tail states of the compositional distribution are similar for all initially mixed compositions, giving rise to very similar PL spectra for all mixed samples after phase segregation. Knowing into which state the material tends to segregate is a vital prerequisite for understanding the phase segregation mechanism. We hope that this knowledge will be used as a starting point to test the segregation models proposed in literature.

ASSOCIATED CONTENT

**Supporting Information**. Experimental & Methods, Full XRD Pattern, Correlating Composition with PL and XRD Peak position, Derivation Stokes-Wilson Formula, Fitting of the XRD Pattern.

AUTHOR INFORMATION

Corresponding Author

Eva L.T. Unger        E-Mail: eva.unger@chemphys.lu.se.        Phone:

ORCID 0000-0002-3343-867x

Ivan G. Scheblykin    E-Mail: ivan.scheblykin@chemphys.lu.se.    Phone:

ORCID 0000-0001-6059-4777

**Notes**

The authors declare no competing financial interest.

ACKNOWLEDGMENT

E. L. U. and K. S. acknowledge financial support from the Swedish Research Council (grants no. 2015-00163 and 2019-05014) and Marie Sklodowska Curie Actions, Cofund, Project INCA 600398 and Nano Lund. E. L. U., A. M. and C. R. also like to acknowledge financial support from the German Federal Ministry of Education and Research (BMBF–NanoMatFutur Project HyPerFORME: 03XP0091). C. R. acknowledges financial support from the HI-SCORE research school of the Helmholtz Association. The work was also supported by the Swedish Research Council grant no. 2016-04433.

# Supporting Information (SI) for:

# Broad Distribution of Local I/Br Ratio in Illuminated Mixed Halide Perovskite Films Revealed by Correlative X-ray Diffraction and Photoluminescence

AUTHOR NAMES Klara Suchan[1], Justus Just[2], Pascal Becker[3,4], Carolin Rehermann[5], Aboma Merdasa[5], Roland Mainz[4], Ivan G. Scheblykin*[1] and Eva L. Unger*[1,5]

AUTHOR ADDRESS [1]Division of Chemical Physics and NanoLund, Lund University, Box 124, 22100 Lund, Sweden, [2]MAX IV Laboratory, Lund University, PO Box 118, SE-22100 Lund, Sweden, [3]Helmholtz-Zentrum Berlin für Materialien und Energie GmbH, Structure and Dynamics of Energy Materials, Hahn-Meitner-Platz 1, 14109 Berlin, Germany, [4]Helmholtz-Zentrum Berlin für Materialien und Energie GmbH, Microstructure and Residual Stress Analysis, Albert-Einstein-Straße 15, D-12489 Berlin, Germany, [5]Helmholtz-Zentrum Berlin für Materialien und Energie GmbH, Young Investigator Group Hybrid Materials Formation and Scaling, HySPRINT Innovation Lab, Kekuléstraße 5, 12489 Berlin, Germany

# SI-Note 1: Experimental methods
## Fabrication

Microscope glass substrates were cleaned in an ultrasonic bath in Mucasol, Aceton and Isopropanol (10 min in each solvent). Subsequently they were dried and cleaned for 15 min in O3-Plasma. Lead(II) iodide and Lead(II) bromide were purchased from TCI (Tokyo Chemical Industry UK Ltd.), MAI and MABr from Greatcell solar and anhydrous N,N-dimethylformamide (DMF) from Sigma Aldrich. All chemicals were used as received. Two solutions were prepared. The iodide-rich perovskite precursor solution was prepared by dissolving 159 mg of MAI and 461 mg of $PbI_2$ into 1 ml anhydrous N,N-dimethylformamide (DMF) . Analogously the bromide-rich perovskite precursor solution was prepared by dissolving 112 mg of MABr and 367 mg of $PbBr_2$ in 1 ml DMF to make 1 M solution with an equimolar ratio of precursor salts. To prepare the mixed halide sample series the two stock solutions were mixed in the appropriate ratios. $60 \mu l$ of the mixed solution were then spin cast onto the substrates at 4000 *RPM* for 30*s*. After spin casting, the samples were annealed on a hotplate at 100 *°C* for 30 min. We studied $MAPb(Br_xI_{1-x})$ samples with compositions $<x>=0.1, 0.2, 0.4, 0.6, 0.8, 0.9$ and 1. To characterize a mixed halide perovskite, we introduce the term mean composition $<x>$. Due to the phase segregation, the mean composition $<x>$ may differ significantly from the nanoscale composition x.

## Measurements

All measurements were performed in a constantly $N_2$ flushed enclosure (shown in Figure 1). The temperature was kept constant at $20 \pm 0.5 °C$ using a peltier element.

*Photoluminescence Spectroscopy*

For in-situ photoluminescence measurements a custom setup was used in which the entire sample was excited with a blue LED (455 nm) at $100 mW/cm^2$ and the PL was detected with a fiber-coupled spectrometer as depicted in Figure 1.

*X-ray Diffraction*

Simultaneous *in-situ* X-ray diffraction measurements were performed at the LIMAX laboratory[1] at BESSY II (Berlin), equipped with a liquid-metal-jet X-ray source, using GaK$_\alpha$-radiation. The samples were measured with temporal resolution of 3 s. For the detection, an area detector (PILATUS3 R 1M) was used. A range of 2θ from 10° to 32° was measured in Bragg-Brentano geometry with the focus on the 200 reflex (2θ≈25°). 1D XRD patterns were obtained by integration over an arc of the 2D diffraction patterns from 5 ° to -5 °azimuthal angle as can be seen in the Figure S1. The full diffraction pattern of all compositions <x> can be seen in Figure S2.

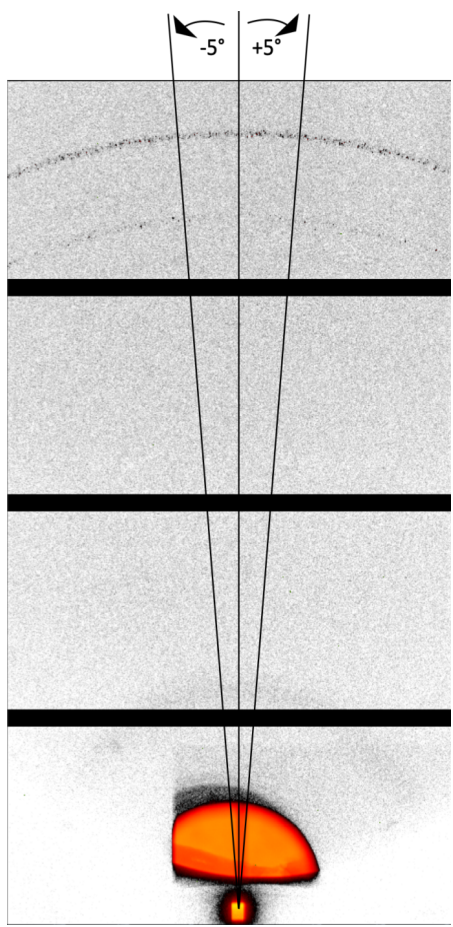

**Figure S1**: The raw x-ray diffraction pattern is shown here exemplarily as obtained from the 2D detector. It was subsequently integrated over an arc between +5° and -5° resulting in a pattern of intensity over diffraction vector length. Using a LaBr6 reference sample this was calibrated and is shown in Figure 1 as a pattern of intensity over the full scattering angle 2Θ.

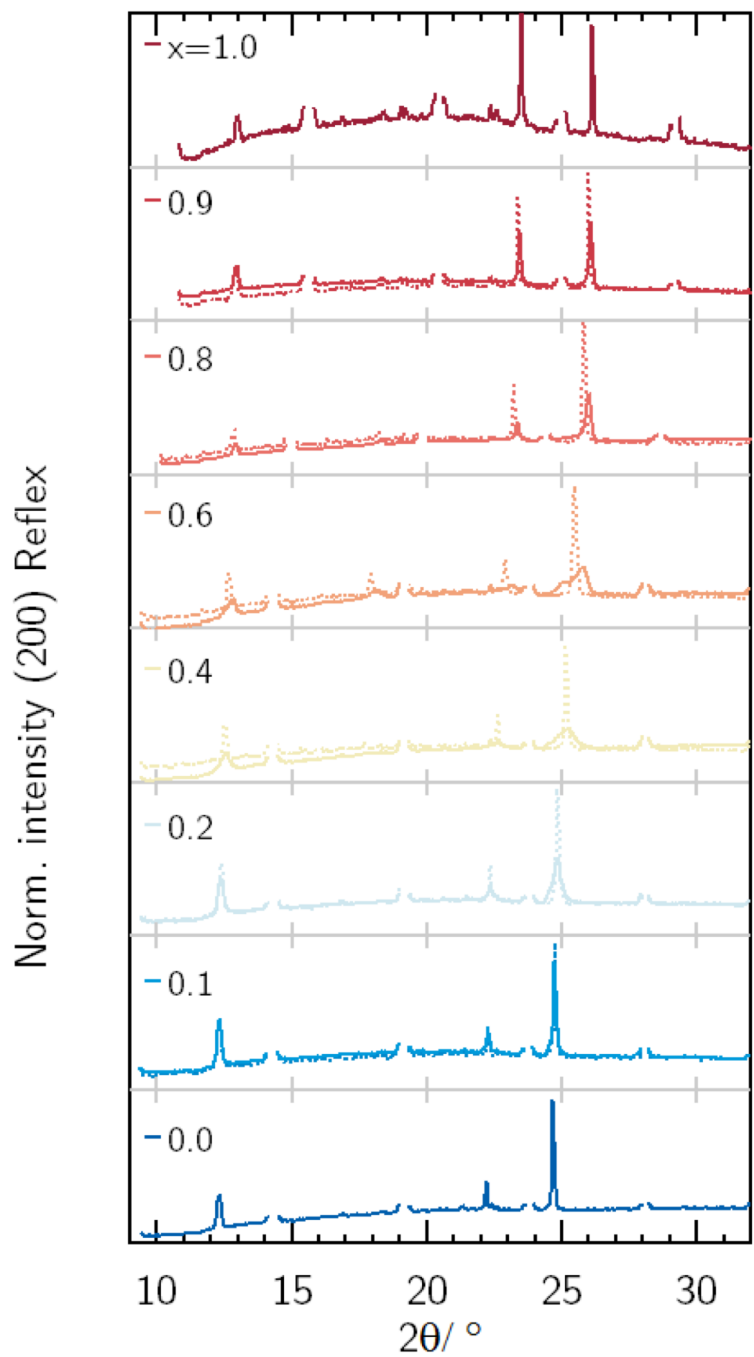

**Figure S2:** Full XRD Pattern from 2θ=10° to 32° for all compositions of MAPb(Br$_x$I$_{1-x}$)$_3$ with x=0 to 1 before illumination (dashed lines) and after illumination (solid lines).

*SI Note2*

*Correlating Composition with PL and XRD Peak position*

In the absence of light, the bandgap, and with it the PL peak energy of MAPb(Br$_x$I$_{1-x}$)$_3$ films, increases for higher Br content due to the smaller size of Br as compared to I as shown previously.[2,3] This increase is almost linear, being well described by modifying Vegard's law[4] by a small additional bowing factor, as shown in Figure S3 b:

$$Eg(x) = Eg_{Br} \cdot x + Eg_I(1-x) - x \cdot (1-x) \cdot b.$$

Assuming, that the PL stems from the band-to-band transition and as prepared samples are sufficiently homogeneous, this correlation between the nominal mean composition <x> and the PL peak energy can be used to estimate the composition of the local phase responsible for PL. We will here refer to such obtained compositions as $x_{PL}$. However, it has to be noted, that from PL measurements, we can obtain compositional information only about states/phases which contribute to the photon emission. Due to relaxation of charge carriers in an inhomogeneous material, these states/phases will contain the lowest part of the electron's density of states of the material. XRD is sensitive to the whole crystalline material in the sample and hence yields information about the overall distribution of compositions in the sample. In absence of light, an increase of the Br content results in an increase in the scattering angle of the mixed MAPb(Br$_x$I$_{1-x}$)$_3$ films in XRD which can also be described by Vegard's law with an additional small bowing factor shown in Figure S3 a, in agreement with previous results[3,5]. Using the relation between scattering angle and Br-content (<x>), the position of the XRD peak can be correlated to the materials composition.

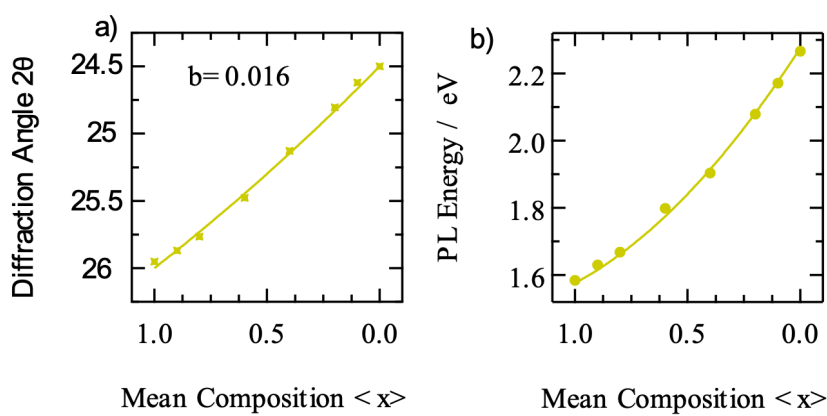

**Figure S3**: a) The peak diffraction angle of the (200) reflex is shown for all compositions together with a fit following Vegard's law with an additional small bowing factor of b=0.016. b) Analogous the photoluminescence peak position for all analyzed mean compositions is shown as dots together with a fit following Vegard's law with a small bowing factor.

*SI Note 3*

*Williamson Hall analysis*

A Williamson Hall (W.H.) analysis of the data for <x>=0 and <x>=0.6, is shown in Fig. S4 together with the LaB$_6$ reference. The LaB$_6$ reference is unstrained and has an average grain size of 10 $\mu$m, such that the peak broadening and thus the slope observed from the W.H. plot can be assumed to give an accurate measure of the instrumental response. The W.H. analysis can be used to detangle the broadening due to the grain size and the lattice distance inhomogeneity. The slight increase of the y-intercept of <x>=0.6 compared to <x>=0 indicates, that there is a slight decrease in the coherence volume, which can be used as a lower

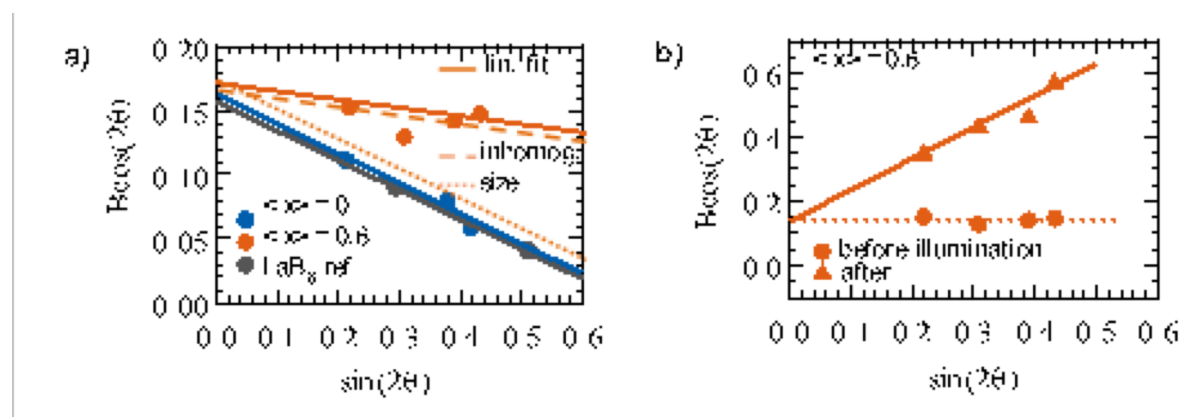

**Figure S4**: a) A Williamson-Hall analysis, where B is the peak broadening, is shown for samples with <x> = 0.6 and <x> = 0 before illumination and for a reference sample (LaB$_6$ powder). The dotted/dashed lines show the limiting cases for the peak broadening expected due to only a decreased size / only increased inhomogeneity, respectively. The main contribution to the increased peak broadening is due to an increase in the inhomogeneity of the lattice distances.

d) Williamson-Hall analysis on the <x> = 0.6 sample before and after illumination.

estimate of the grain size. This however, cannot fully explain the observed broadening as indicated by the dotted line. Instead, the W.H. analysis for <x>=0.6 indicates, that the increased peak width originates primarily from a lattice distance inhomogeneity in the mixed crystal.

A W. H. analysis can analogously be used to rationalize the observed peak broadening under illumination. The results for the <x>=0.6 composition before and after illumination are presented in Fig. S4b. The agreement of the y-intercept for the sample before and after illumination indicates, that the change in coherence volume is negligible and thus the phase-segregation into low- and high- energy domains does not result in a significant decrease of grain size. Compared to the unilluminated sample, the W. H. analysis shows a further increase of the slope after illumination which has previously been assigned to an increase of the inhomogeneous micro strain.[9,25] However, we would like to emphasize here, that any variation, δd, of the lattice plane distance, d, independent of its origin, will result in the same peak broadening, δθ, according to the Stokes Wilson Formula:

$$n\lambda = 2d\sin\theta \quad \& \quad n\lambda = 2(d - \delta d)\sin(\theta - \delta\theta) \quad \rightarrow \quad \delta\theta = -\frac{\delta d}{d}\tan\theta$$

From this equation, the peak broadening due to compositional inhomogeneity is virtually indistinguishable from the strain-related broadening.

The Stokes Wilson Formula is used in a Williamson Hall analysis to relate peak broadening in θ to inhomogeneous strain. However, from the derivation of the Stokes Wilson Formula it becomes clear, that any inhomogeneity of the lattice plane distances d, results in the same tan θ dependence of the peak broadening in θ. This is independent of the origin and the nature of the lattice distance inhomogeneity.

$\lambda n = 2d\sin\theta$ as well as $\lambda n = 2(d - \delta d)\sin(\theta - \delta\theta)$

$d\sin(\theta) = (d - \delta d)\sin(\theta - \delta\theta)$

$d\sin(\theta) = (d - \delta d)(\sin\theta - \cos\theta\,\delta\theta)$

$d\cos\theta\,\delta\theta = -\delta d\sin\theta$

$\delta\theta = -\frac{\delta d}{d}\tan\theta$

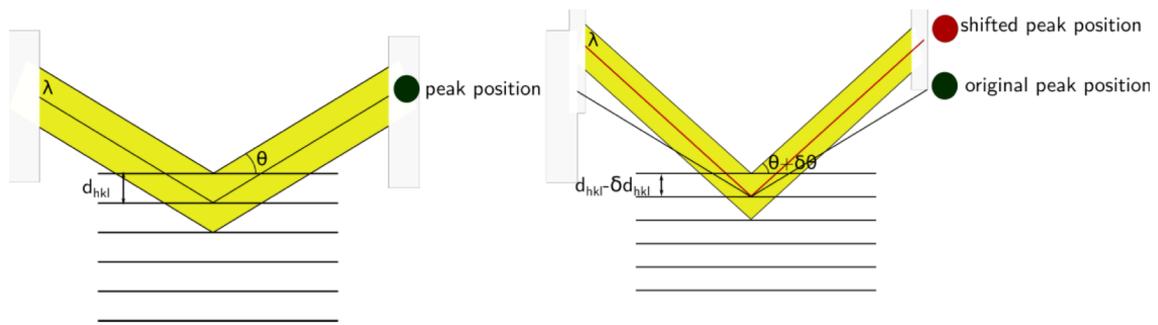

**Figure S5:** Sketch of the diffraction of an X-ray beam with wavelength λ on a crystal with lattice plane distance $d_{hkl}$ and a crystal with reduced lattice plane distance of $d_{hkl}-\delta_{dhkl}$, resulting in two different diffraction angles of θ and θ-δθ.

Thus, a decrease in d, due to a higher ratio of Br incorporated within the lattice, is expected to create an increase in θ, indistinguishable to compressive strain.

*SI Note 4*

*Calculation of the arithmetic mean of all points on the x axis weighted by their individual intensity (center of mass)*

The center of mass of the XRD patterns can be used to investigate any change in net strain. Both compressive and tensile strain can occur in a material. Compressive strain leads to a shift of the XRD peak to higher angles, while tensile strain leads to a shift to lower angles.

Both compressive strain and tensile strain may compensate each other, such that a shift of the entire XRD peak corresponds to a change in the average strain in the sample. To calculate the peak position of the entire peak independent of peak shape, an arithmetic mean of all points on the horizontal axis weighted by their individual intensity can be calculated according to:

$$COM = \frac{\sum_{m=i}^{f} x \cdot y}{\sum_{m=i}^{f} y} \qquad (1)$$

This will further be referred to as center of mass (COM) of the peak.

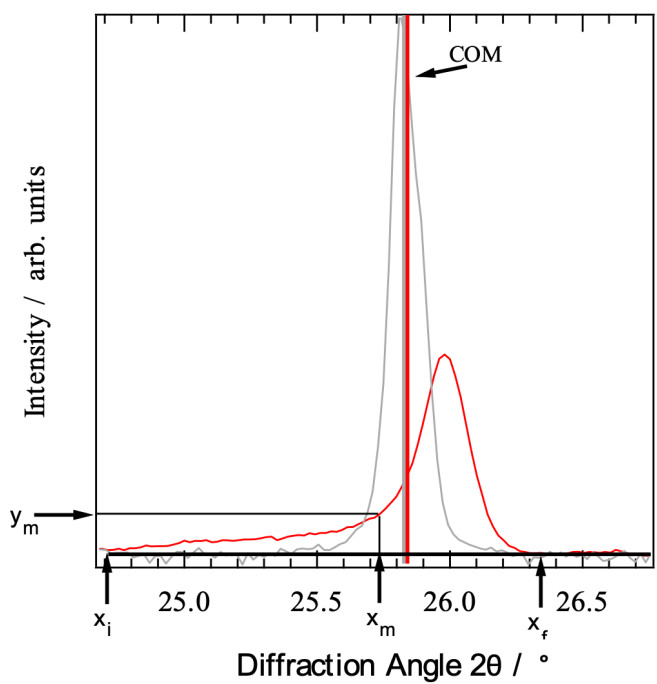

**Figure S6:** Diffraction pattern of the 200 reflex of sample <x>=0.8 before and after illumination(grey/red). The COM of each is represented by a solid vertical line in grey and red, respectively.

Even if tensile and compressive strain compensate each other, such that the net strain is zero, the XRD peak may get affected by the strain. If some parts of the material are under compressive and others under tensile stress, part of the XRD peak is shifted to higher angles and another part of the peak to lower angles, leading to a broadening of the peak. Such a broadening due to strain is typically of Gaussian shape.

*SI Note 5*

*Fitting of the X-ray diffraction pattern:*

Before illumination, the 200 peak of the diffraction pattern for each composition can be fitted with a Pseudo Voigt (PSV) with a unique peak width, (Figure S4). The such obtained peak width is small for samples with compositions close to <x>=0 and <x>=1 and broadens for compositions close to <x>=0.5 as (Figure 3a). The phase segregated diffraction patterns are additionally broadened compared to the non-illuminated film, which can be seen in Figure 4a.

In the case of <x>=0.6, two peaks are visible in the diffraction pattern. In Fig. S6 the 200 peak from the sample <x>=0.6 is shown together with two PSV peaks of composition x=0.3 and x=0.7. The peak width of the PSV is kept the same as for the unilluminated samples of <x>=0.3 and <x>=0.7. This clearly illustrates that next to a peak split an additional broadening of each peak is occurring. If one assumes a split into two compositions and an additional peak broadening stemming from micro strain, the peak should resemble a convolution of the original peak shape (Pseudo Voigt) and a Gaussian. While this describes the high energy peak well, the fit of the low energy peak differs from the measured data (Figure S5 b). This results in a fitted average composition <x> as obtained from the fit of 0.7, which differs from the actual average composition of <x>=0.6. Hence, assuming a split into two compositions with a strain related peak broadening does not accurately describe the pattern.

The pattern can neither be sufficiently described by a PSV function with an additional gaussian broadening, nor by Gaussian lineshape. Instead the patterns were fitted by a linear combination of 20 different compositions from x=0 to x=1, as shown in Figure 4c. We assume that the peak width of the phase segregated samples depends on the local composition x in the same way as it is dependent on <x> for the samples before illumination, as shown in Figure 3a. This is a reasonable assumption, as the contributions from micro strain and size were shown to be negligible. The fit is constraint to ensure that the average composition <x> stays constant before and after illumination.

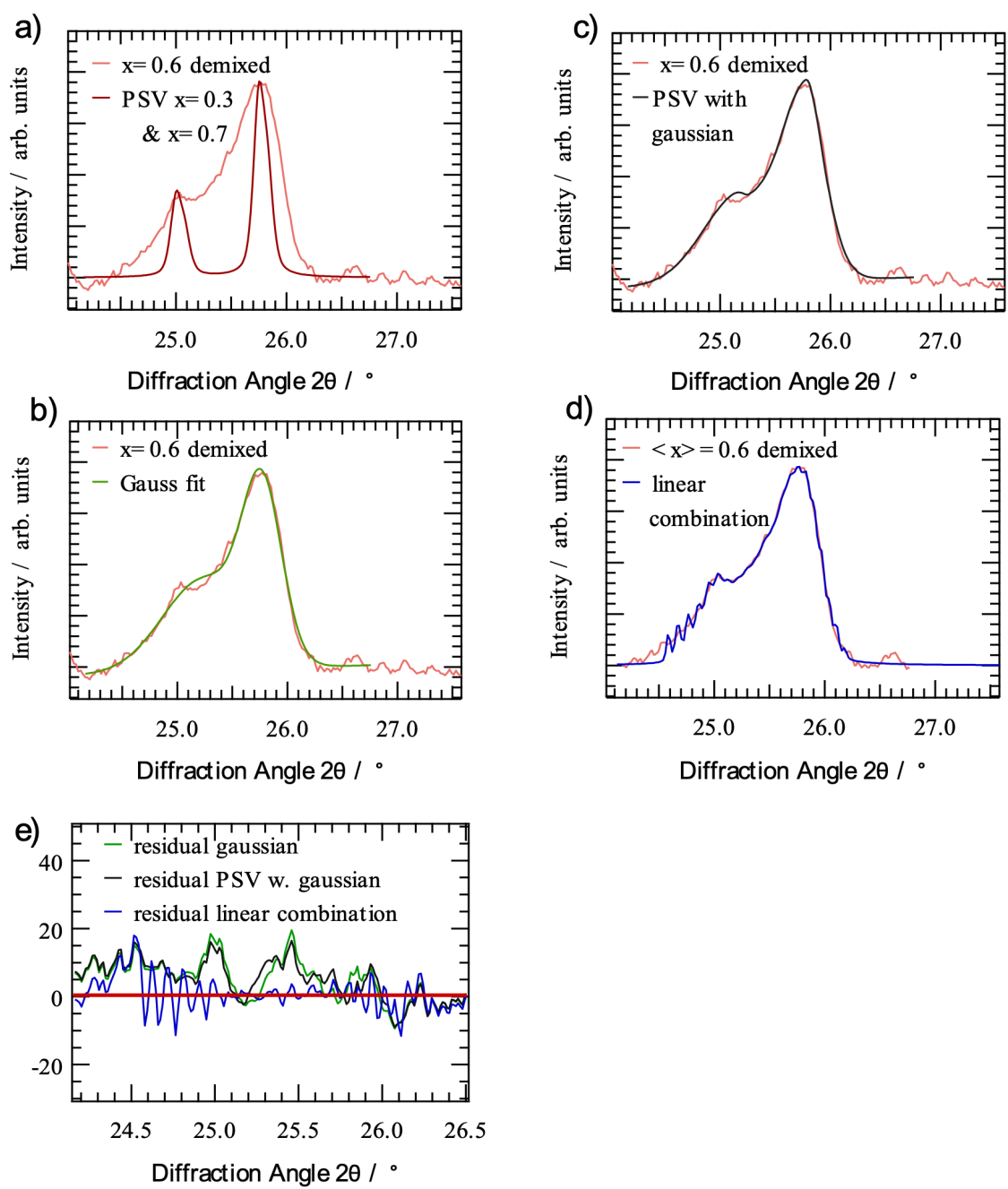

**Figure S7:** Fitting the X-ray diffraction pattern measured from the sample with composition <x>=0.6. a) Fitting attempt with 2 Pseudo Voigt (PSV) peaks with the same peak width as found for the unperturbed samples. (Figure 3). The peak position resembles compositions of x=0.3 and 0.7. b) Fit with 2 PSV peaks with an additional gaussian broadening. c) Fit with 2 Gaussians with unrestricted peak width. d) Fit with a linear combination of 20 PSV peaks as described above. e) shows a comparison of the residuals for the fits shown in b,c and d.